\documentclass{aastex61}
\usepackage{lineno}

\newcommand\aastex{AAS\TeX}

\shorttitle{\aastex\ Follow up GW170104}
\shortauthors{Fermi GBM-LAT team}

\begin{document}

\title{\emph{Fermi} observations of the LIGO event GW170104}

\author{On behalf of the \emph{Fermi}-GBM Collaboration}


\author{the \emph{Fermi}-LAT Collaboration}

\begin{abstract}
We present the \emph{Fermi} Gamma-ray Burst Monitor (GBM) and Large Area Telescope (LAT) observations of the LIGO binary black hole merger (BBH) event GW170104.  No candidate electromagnetic counterparts was detected by either GBM or LAT. A detailed analysis of the GBM and LAT data over timescales from seconds to days covering the LIGO localization region is presented. The resulting flux upper bound from the GBM is (5.2--9.4)$\times$10$^{-7}$ erg cm$^{-2}$ s$^{-1}$ in the 10-1000 keV range and from the LAT is (0.2--13)$\times$10$^{-9}$ erg cm$^{-2}$ s$^{-1}$ in the 0.1--1 GeV range. We also describe the improvements to our automated pipelines and analysis techniques for searching for and characterizing the potential electromagnetic counterparts for future gravitational wave events from Advanced LIGO/VIRGO.

\end{abstract}
\keywords{gravitational waves, gamma rays:  general, methods: observation}

\section{Introduction} \label{sec:intro}
The detection of gravitational waves (GWs) has opened a new window to the cosmos and paved the way to a new era of multi-messenger astronomy. The first detected signals by Advanced LIGO, GW150914 and GW151226 ~\citep{Abbott2016gw150914, Abbott2016gw151226} are compatible with the coalescence of high-mass binary black holes (BBHs). The detection of a third BBH merger, GW170104~\citep{GW170104} (t$_{GW}$ = 2017-01-04 10:11:59 UTC), with component masses of $\sim 31 \ \rm M_\Sun$ and $\sim 19 \ \rm M_\Sun$, adds to the growing sample BBH merger events.  The component masses and the post-merger mass ($\sim 49 \ \rm M_\sun$) of GW170104 falls between those of GW151226 and GW150914, while the estimated distance of $\sim 880$ Mpc ($z\approx 0.18$) is roughly twice as far compared to the previous detections.  GW binary coalescence events are also predicted to result from the mergers of other compact object binary systems such as neutron star-black hole (NS-BH)
and neutron star-neutron star (NS-NS). Short gamma-ray bursts (sGRBs) are thought to be associated with such systems~\citep{Eichler89,Narayan92,Lee07}, which strongly motivate search for electromagnetic (EM) counterparts to GW events.


The {\it Fermi} Gamma-ray Burst Monitor~\citep[GBM,][]{Meegan09} is the most prolific detector of sGRBs~\citep{Bhat16}($\sim$40 per year); however, it localizes these sources with uncertainties of the order of a few degrees, making the follow-up by some instruments at other wavelengths challenging. The {\it Fermi} Large Area Telescope~\citep[LAT,][]{Atwood09} has a lower detection rate for sGRB  but can provide 0.2--0.3$^{\circ}$ localizations. In the case of a detection of an EM counterpart, the LAT could substantially reduce the localization uncertainty, facilitating follow-up at other wavelengths. The two instruments on-board the {\it Fermi} satellite are complementary and are uniquely capable of providing  all-sky  observations  from  hard X-rays to  high-energy gamma rays in normal survey  operations. Together, they cover the  entire  localization probability maps of gravitational wave events~\citep{Ackermann16,Racusin17} within hours of their detections.

\section{Observations and data analysis}

\subsection{GBM}
The {\it Fermi} GBM has a collection of 12 sodium iodide (NaI) detectors and 2 bismuth germanate (BGO) detectors sensitive over an energy range of 8 keV--40 MeV with 128-channel spectral resolution~\citep{Meegan09}. The detectors observe the full un-occulted sky ($\sim 70\%$) and operate continuously except during passage of {\it Fermi} through the South Atlantic Anomaly (SAA), so that the average duty cycle for observing a given sky location is $\sim$60\%.  Since late 2012, GBM has provided Continuous Time-Tagged Event data (CTTE) with resolution down to 2~$\mu$s, which, together with the large spectral range and sky coverage, makes GBM the most prolific detector currently in operation for prompt GRB observations.

\subsubsection{GBM Signal Searches}
An automatic, ground-based search of the GBM CTTE data for sGRBs is currently running promptly after the data are downlinked.  This search, which we refer to here as the un-targeted search, is similar to the flight software algorithms that GBM uses for triggering on GRBs, which look for simultaneous excesses in two GBM NaI detectors. This search currently rebins the CTTE data into time series with 18 bin widths from 64 ms to 30.72~s and four phase offsets for each bin width.  These time series are created for four different energy ranges that were optimized for finding weak triggered sGRBs: nominally 30--500 keV, 50--500 keV, 100--500 keV and 100 keV--1 MeV.  The background estimation is performed via cubic spline fitting and intervals of data are removed from consideration if the background fit deviates significantly from the observed background.  While the in-orbit triggering algorithms require signals to reach 4.5$\sigma$ in two separate detectors, the un-targeted search flags any signals that exceed 2.5$\sigma$ in one detector and 1.25$\sigma$ in another.  A pre-trials p-value is calculated for the flagged candidate and compared against a pre-determined p-value threshold to determine whether a candidate signal has been found.  Archival CTTE data have been searched starting from January 2013, finding $\approx$80 candidate sGRBs  per year, in addition to the 40 sGRBs found by in-orbit triggering\footnote{\url{https://gammaray.nsstc.nasa.gov/gbm/science/sgrb_search.html}}.

The GBM targeted search pipeline~\citep{Blackburn15}, which operates using a seed time and optionally a seed sky map, was updated in preparation for the second observing run of Advanced LIGO (O2) to produce an overall lower False Alarm Rate (FAR) and to be more sensitive to sGRB-like signals~\citep{Goldstein16}.  The improvements implemented include replacing the `hard' spectral template with a more physically and observationally representative cut-off power-law spectrum, increasing the data temporal resolution, improving the background 
estimation, and utilizing localization information in the ranking statistic.

\subsubsection{GBM Results}
No on-board GBM trigger was recorded close in time to GW170104.  The nearest triggers were more than thirteen hours preceding and more than 10 hours following GW170104, and both of these triggers were due to steep non-astrophysical rate increases near the boundary of the SAA just before the detectors were turned off for passage. The un-targeted search did not find any un-triggered sGRB candidate during January 4.  The nearest candidate preceding GW170104 was more than 15 hours previous on January 3, and the nearest candidate following GW170104 was about 8 days later on January 12.

The targeted search was run over a 1 minute window centered at the LIGO trigger time of GW170104, searching for any signal on timescales from 0.256 s, increasing by factors of two up to 8.192 s (6 timescales).  As reported in~\citet{Burns17}, no credible significant candidate was found.  The most significant result from the search was on the longest timescale with the `soft' spectral template, which the search found to be marginally consistent with part of the LIGO sky map.  Further inspection of the count rate lightcurves for each detector reveal a slow-rising bump that begins $\sim$100 s before the LIGO trigger and has a duration of $\sim$300 s.  This bump, which is strongest in the GBM NaI channel 1 (12--25 keV) and does not appear above channel 2 (50 keV), is consistent with a low-level increase in magnetospheric activity as {\it Fermi} proceeds through its orbit. 

\begin{figure}
	\begin{center}
		\includegraphics[scale=0.4]{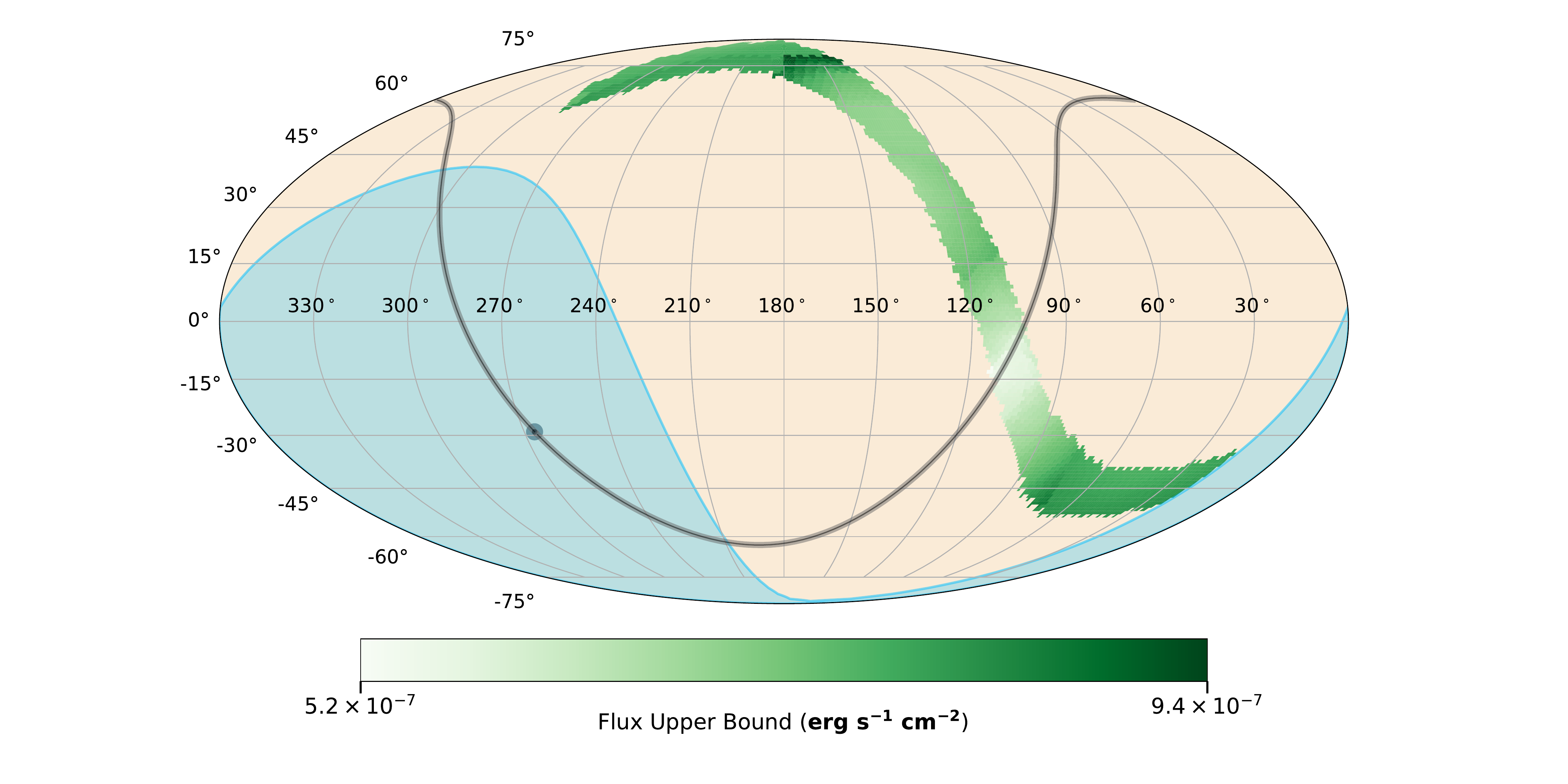}
	\end{center}
\caption{The 1 s flux upper bound by GBM covering the 99\% credible region of the LIGO sky map.  The Earth occultation region is shown in blue, and the Galactic Plane is shown as the gray line with the Galactic center demarcated with a gray circle.
\label{GBMUpperLimit}}
\end{figure}

Using GBM, we can set an upper bound on impulsive flux expected from a sGRB-like signal.  The current technique is to estimate the background in the same way that the targeted search does (un-binned Poisson maximum likelihood over a sliding window) and calculate the 3$\sigma$ upper bound in the count rate as 3 times the standard deviation of the background count rate.  The count rate upper bound can then be converted to a flux upper bound by assuming a photon model, folding it through the detector response and fitting the amplitude to determine the associated flux upper bound.  We note that this technique produces a conservative upper limit because it utilizes a single GBM detector rather than all detectors over the entire LIGO map, which would be significantly more computationally challenging.  We evaluate the limits in this way over the full 99\% credible region of the visible LIGO sky map by splitting up the map into a set of equally spaced regions on the sky and considering the GBM NaI detector with the smallest pointing angle to the center of each region.  Performing this technique over the targeted search window, we estimate the 1 s averaged flux upper bound assuming a cut-off power law with $E_{\rm peak} = 566$ keV and photon index of $-$0.42, which represents the peak density of the spectral parameter distribution for GBM-triggered sGRBs\footnote{\url{https://heasarc.gsfc.nasa.gov/W3Browse/fermi/fermigbrst.html}}.  The resulting spatially-dependent upper bound is shown in Figure~\ref{GBMUpperLimit} and ranges from $(5.2-9.4)\times10^{-7}$ erg $\rm s^{-1}$ $\rm cm^{-2}$ in the 10--1000 keV range. These upper bounds are below 90--99\% of GBM-triggered sGRBs with a similar spectrum to the model used for the upper bound estimation. Based on the distance of GW170104, $D_L\approx880$ Mpc, $z\approx0.18$, the flux upper bound translates to a K-corrected luminosity upper bound range of $(5.5-9.9) \times 10^{49}$ erg s$^{-1}$ in the approximately bolometric 1--10$^4$ keV energy range typically reported for GRBs.

In addition to upper bounds for impulsive emission, we searched for any long-lasting or persistent emission related to GW170104 using the Earth Occultation Technique~\citep[EOT;][]{Wilson-Hodge12}.  The search was performed for a two-day interval centered on the GW170104 trigger time for any sources within the 99\% LIGO localization credible region. No new sources were detected during this interval and no significant change in flux was recorded for any of the sources that GBM typically observes with the EOT.  A search was also performed for a 30 day period after the trigger, and again no new source was detected.

\subsection{LAT}
The LAT~\citep{Atwood09} is a pair conversion telescope comprising a 4$\times$4 array of silicon strip trackers and tungsten converts together with cesium iodide (CsI) calorimeters covered by a  segmented  anticoincidence  detector  to  reject charged-particle  background  events. The  LAT is sensitive in the  energy range from 20~MeV  to  more than 300~GeV with a field of view (FoV) of 2.4~sr, observing the  entire  sky  every  two  orbits  (3~hours)  by rocking north and south about the orbital plane on alternate orbits~\citep{Atwood09}. The LAT detects about 15 GRBs per year above 100 MeV, $\sim$2--3 of which are sGRBs with localization precisions on the order of $\sim$10 arcmin~\citep{Vianello15}. This latter class of GRBs, when detected above 100 MeV,  has a substantially longer duration with respect to their keV--MeV emission and is thought to be related to the afterglow phase. The LAT is the only instrument that has detected and localized a sGRB during its afterglow phase starting from the GBM localization~\citep{Ackermann10}. Furthermore, the LAT can substantially reduce the localization uncertainties with respect to GBM, aiding follow-up at other wavelengths.

\subsubsection{The LAT signal searches}
Several automatic programs search for transient events over a wide range of timescales in the LAT data. The Fermi automatic science processing (ASP) pipeline and the Fermi All-sky Variability Analysis (FAVA)\footnote{\url{https://fermi.gsfc.nasa.gov/ssc/data/access/lat/FAVA/}} are used to search for excess emission on hour-to-week long timescales. The ASP pipeline performs a blind search for sources on all-sky count maps constructed from the event data acquired on 6 and 24 hr timescales. Candidate flaring sources are then fit using a standard likelihood analysis with the model including nearby known sources and the Galactic and isotropic diffuse contributions. These candidate sources are then characterized and matched to known sources, allowing for the detection of flaring cataloged sources as well as new unassociated sources.  FAVA is a blind photometric analysis in which a grid of regions covering the entire sky is searched over 24 hr and 1 week timescales for deviations from the expected flux based on the observed long-term mission-averaged emission \citep{Ackermann13}. This allows the FAVA search to be independent of any model of the gamma-ray sky, and therefore complementary to the standard likelihood analyses.

As described in \citet{Ackermann16,Racusin17} and, more specifically, in \citet{Vianello17}, we have designed specific methods for the follow-up searches in the LAT data of EM counterparts to GW events. The fixed time window search and the adaptive time window search are two of these custom search methods that we run on the LIGO localization maps. In the first case, we search for high-energy gamma-ray emission on a set of fixed time windows ($\approx$~10ks long) around the LIGO trigger. For each time window, we select all the pixels\footnote{In our analysis, the average separation between pixels is $\approx$~0.5$^{o}$} that were observable by the LAT within the 90\% containment of the LIGO localization map and perform an independent likelihood analysis in an 8$^{\circ}$ radius region of interest (ROI). In the latter method, we optimize the time window for the analysis based on when the pixel becomes observable by the LAT. For each pixel we further select only the interval that contains the GW trigger time, or the one immediately after (if the center of the pixel was outside of the LAT FOV at the GW trigger time). Once we have defined these adaptive time windows, we perform an independent likelihood analysis for each pixel. This  analysis  is  therefore  optimized  for  the  assumption that the source emitted gamma rays at the time of the GW event. In both of these analysis methods we use the Pass 8 \texttt{P8\_TRANSIENT010E\_V6} events class and the corresponding instrument response functions and perform the analysis from 0.1 to 1 GeV. Our  analysis  is  based  on  the  standard  unbinned  maximum  likelihood  technique  used  for LAT data analysis.  We include in our baseline likelihood  model  all  sources  (point-like  and  extended)  from  the  3FGL catalog of LAT sources~\citep{Acero2015},   as  well  as  the  Galactic and isotropic diffuse templates provided in~\citet{Acero2016}. We  employ  a  likelihood-ratio  test~\citep{Neyman1928} to quantify whether the existence of a new source is statistically warranted.  In doing so, we form a test statistic (TS) described in~\citet{Vianello17} to reject a null hypothesis of no signal associated with the GW trigger.
As is standard for LAT analysis, we choose to reject the null hypothesis when the TS is greater than 25, roughly equivalent to a 5$\sigma$ rejection criterion for a single search.

In preparation for the LIGO O2 science runs, we have implemented several improvements to our GW follow-up searches. These include an analysis pipeline to automatically process LAT data every time a LIGO/VIRGO Gamma-ray Coordinates Network (GCN) is received, running both the fixed time window search and the adaptive time window search methods.

\subsubsection{LAT observations of GW170104}
The LAT was favorably oriented toward GW170104 at the time of the trigger, t$_{GW}$, covering ~55\% of the LIGO localization map at t$_{GW}$, and within 5 ks from t$_{GW}$ the LAT had observed 100\% of the LIGO probability map. The LAT then continued to observe the entire LIGO localization region throughout normal sky-survey operations in the following days. We performed a fixed time window search from t$_{GW}$ to t$_{GW}$+10 ks and found no credible candidate counterparts. We thus performed the upper bound computation described in~\citet{Vianello17} to evaluate the 95\% upper bound in the 0.1--1 GeV range for the fixed time interval and found a value of F$_{ul,95}$= 3$\times$10$^{-10}$ erg cm$^{-2}$ s$^{-1}$.

\begin{figure}
	\begin{center}
     \includegraphics[scale=0.4]{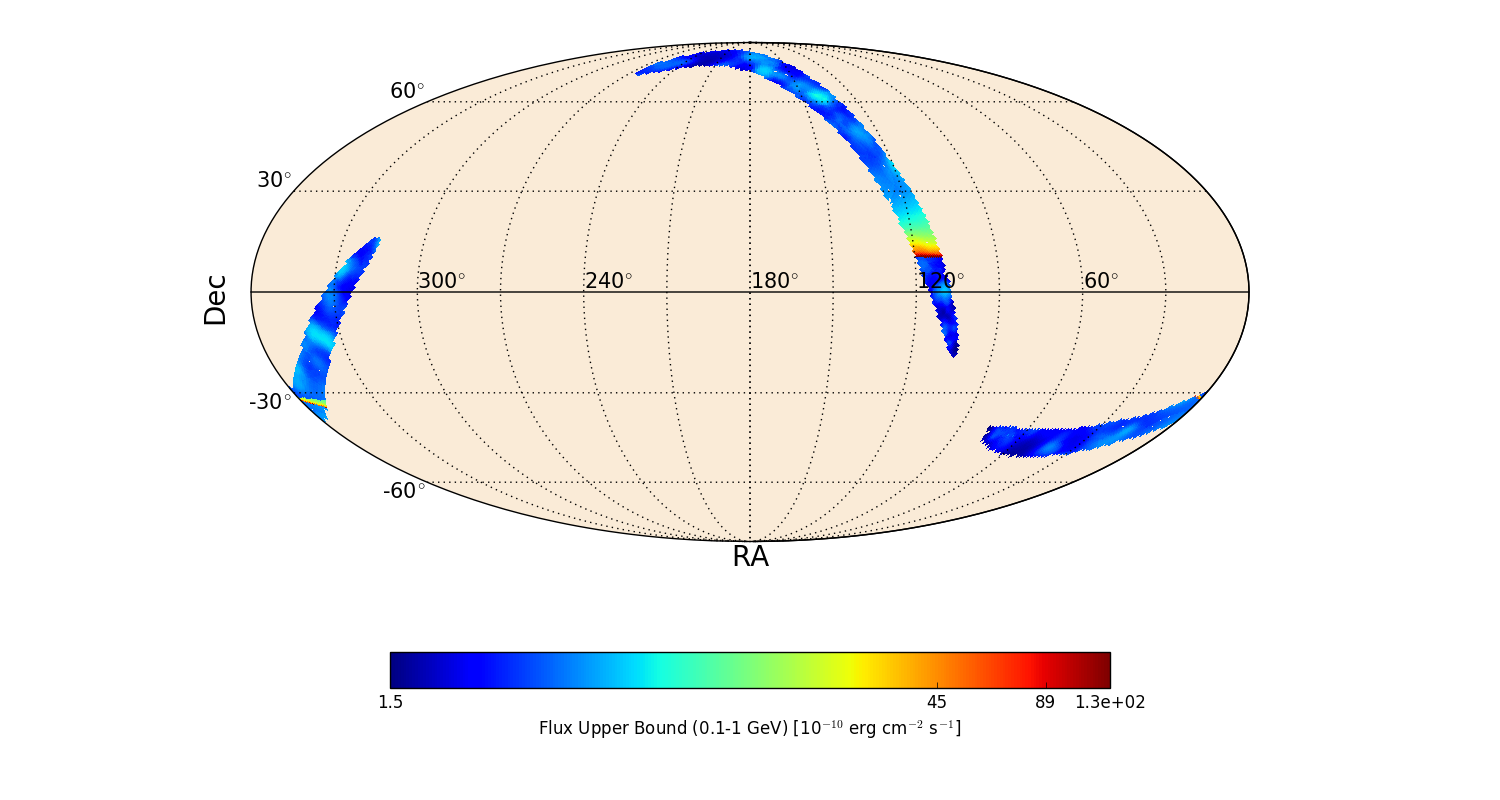}	
	 \includegraphics[scale=0.3]{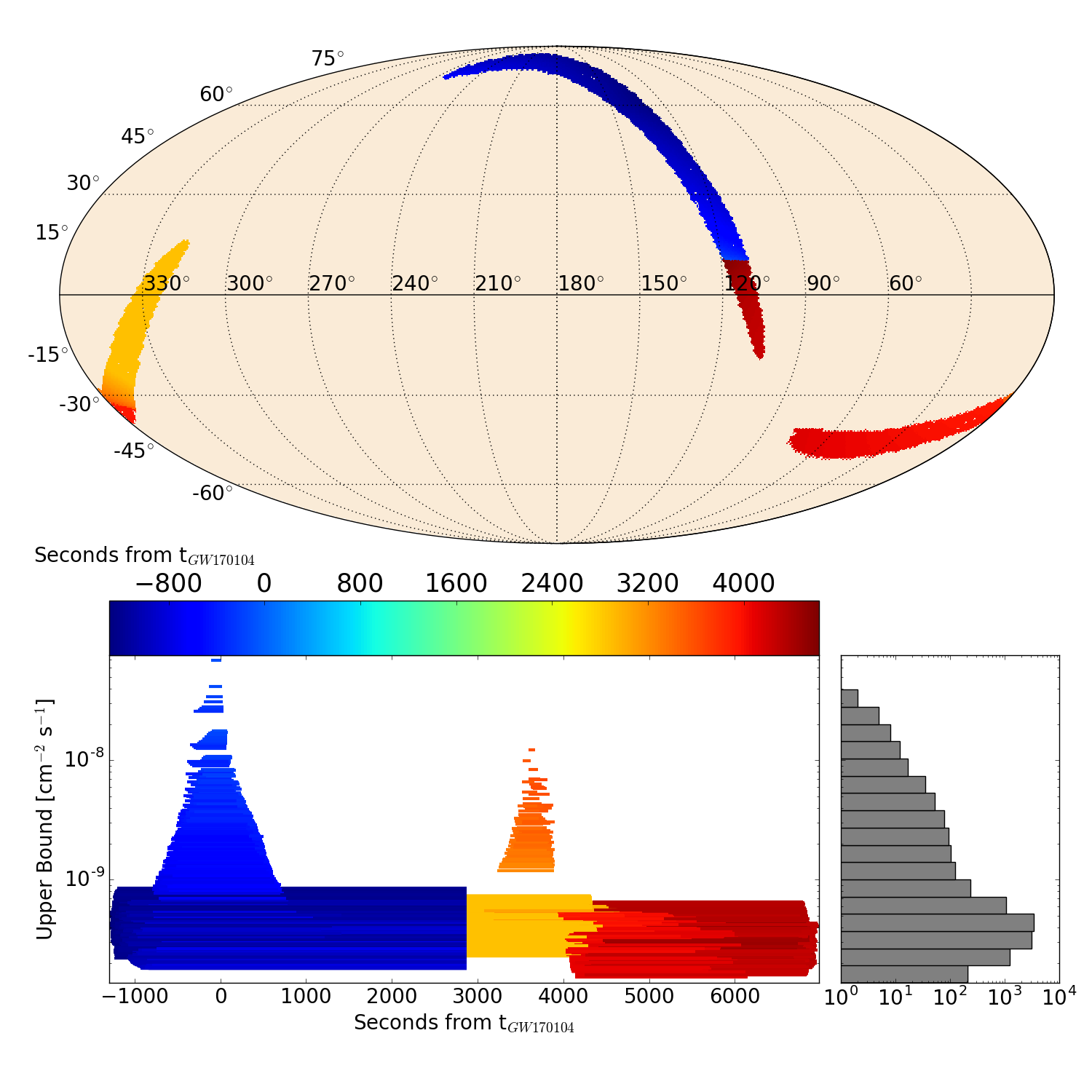}
	\end{center}
\caption{The adaptive time interval analysis for GW170104. Flux upper bound map during the time adaptive analysis (top), the entry time into the LAT FOV relative to the t$_{GW}$ for each pixel within the LIGO localization contour (middle), and the upper bound light curves for each ROI (bottom). The horizontal bars in the bottom panel correspond to the values of the LAT upper bounds, and their position along the time-axis coincides with the interval of time used in the analysis. The color of each bar indicates the time when the 8-degrees ROI entered the LAT FOV, and matches the color of the pixel in the middle panel. The horizontal histogram displays the distribution of upper bounds.}
\label{LATUpperLimitsAdaptive}
\end{figure}

Also the adaptive time window analysis did not yield any significant excesses and no new sources were detected above a 5$\sigma$ detection threshold. The flux upper bounds in the 0.1--1 GeV range for the portion of the LIGO localization contour containing 90\% of the probability during the adaptive time window are shown in Figure~\ref{LATUpperLimitsAdaptive}. The values for the flux upper bounds from this analysis range from 0.2--13$\times$10$^{-9}$ erg cm$^{-2}$s$^{-1}$.

Given that we do not know the nature of the potential EM counterpart to a BBH merger, we also examined the ASP products during the 6 and 24 hr intervals containing t$_{GW}$. We also performed a dedicated FAVA search over a 24 hr and 1 week intervals centered on t$_{GW}$. The two searches found two flaring sources with significance above 5$\sigma$ within the LIGO localization map.  Follow-up likelihood analyses of both sources found that their positions were consistent with flat spectrum radio quasars (S5 1044+71 and OK 630) which have shown previous evidence of flaring activity. No new unassociated flaring sources were detected within the LIGO localization contour.

\section{Summary}

It is unclear how BBH mergers can be followed by BH-accretion disk systems that are conventionally invoked \citep[e.g.][]{2014ARA&A..52...43B} to channel the gravitational energy into EM radiation. For GW150914 \citet{Connaughton16} reported a possible counterpart, which increased interest in scenarios by which BBH mergers may yield significant gamma-ray emission. However, all current models suffer from non-negligible inconsistencies \citep[e.g.][]{Lyutikov16} and for this reason even upper bounds like the ones presented here contribute important information to the puzzle. Some models \citep{Loeb16,Woosley16}, involve the presence of a common envelope or a circumbinary disk and have been also used to explain possible neutrino emission \citep{Janiuk17,2016PhRvD..93l3011M}. According to \citet{Perna16}, a sGRB might result from the BBH merger triggering the accretion of a disk around one of the BHs in the final seconds before the coalescence, but \citet{Kimura17} find it is difficult to have the disk survive down to seconds before the coalescence as required. An efficient way to power EM emission is the Blandford-Znajek~\citep{Blandford77} process of electromagnetic extraction of mechanical energy \citep{Li16,Veres16}. \citet{Lyutikov16} however pointed out that the EM luminosity associated with GW150914 would require magnetic fields of $\sim$10$^{12}$ G, astrophysically  implausible for an environment around a BH.

Based on the general arguments listed in \citet{Racusin17}, non-detection of an EM counterpart for GW170104 does not rule out the association between GW150914 and GW150914-GBM. These arguments are: a.) The source may have been occulted by the Earth from {\it Fermi} at the time of trigger~\citep[$\sim$18\% of the LIGO map was occulted;][]{Burns17}. b.) The GBM background rates are higher for GW170104 (average 113 counts s$^{-1}$ per detector, or 11\%, higher compared to GW150914-GBM).  Note that the estimated flux for the $\sim 1$ s long signal GW150914-GBM~\citep[$\sim2\times10^{-7}$ erg s$^{-1}$ cm$^{-2}$;][]{Connaughton16} is below the GBM flux upper limit we estimate here, indicating that it is unlikely that GW150914-GBM could be detected with this increased background rate. c.) A jet, if present, was pointing away from us. d.) The EM emission associated with GW170104 was dimmer than for GW150914 (e.g., related to the lower BH masses and the larger source distance).

GW170104 is the third BBH merger detected by Advanced LIGO. While no conventional model predicts EM counterparts for these sources, our observations are important in two respects: a.) We improve and test the search algorithms in anticipation of GW events where we do expect an EM counterpart (e.g., a binary merger with at least one NS component). b.) We provide flux upper bound constraints for the unexpected scenario where BBH mergers are accompanied by gamma rays.

The \textit{Fermi}-GBM collaboration acknowledges support for GBM development, operations and data analysis from NASA in the United States and BMWi/DLR in Germany. The \textit{Fermi}-LAT Collaboration acknowledges support for LAT development, operation and data analysis from NASA and DOE (United States), CEA/Irfu and IN2P3/CNRS (France), ASI and INFN (Italy), MEXT, KEK, and JAXA (Japan), and the K.A.~Wallenberg Foundation, the Swedish Research Council and the National Space Board (Sweden). Science analysis support in the operations phase from INAF (Italy) and CNES (France) is also gratefully acknowledged. This work performed in part under DOE Contract DE-AC02-76SF00515.


\end{document}